\documentclass{svjour2}

\usepackage{graphicx}  
\usepackage{latexsym}
\usepackage{amssymb}

\def\beq{\begin{equation}}
\def\eeq{\end{equation}}

\journalname{General Relativity and Gravitation}

\begin{document}

\title{Scattering by a Schwarzschild black hole of particles undergoing drag force effects}

\author{Donato Bini \and 
Andrea Geralico
}

\institute{Donato Bini 
\at
Istituto per le Applicazioni del Calcolo ``M. Picone,'' CNR, Via dei Taurini 19, I-00185, Rome, Italy\\
\email{donato.bini@gmail.com} 
\and
Andrea Geralico 
\at
Istituto per le Applicazioni del Calcolo ``M. Picone,'' CNR, Via dei Taurini 19, I-00185, Rome, Italy\\
\email{geralico@icra.it}       
}

\date{Received: date / Accepted: date / Version: \today}

\maketitle

\begin{abstract}
The scattering of massive particles by a Schwarzschild black hole also undergoing a drag force is considered.
The latter is modeled as a viscous force acting on the orbital plane, with components proportional to the associated particle 4-velocity components. The energy and angular momentum losses as well as the dependence of the hyperbolic scattering angle on the strength of the drag are investigated in situations where strong field effects cause large deflections.
\end{abstract}

\section{Introduction}

Interactions or collisions of two black holes occupy a central role in the recent literature of binary systems, especially because of the associated emission of gravitational radiation which is expected to be in the range of detectability of most of the Earth interferometers, like LIGO \cite{ligo}, VIRGO \cite{virgo}, etc., as in the event GW150914 \cite{Abbott:2016blz}.  
In this context, a number of works has been produced either concerning the study of strongly inelastic collisions in the framework of full numerical relativity 
\cite{Pretorius:2007jn,Shibata:2008rq,Sperhake:2008ga,Sperhake:2009jz,Witek:2010xi,Sperhake:2012me}
(leading to a prompt merger of the two black holes), or focusing on hyperbolic-like elastic or quasi-elastic 
scattering by a combined use of analytical and numerical approaches as in Ref. \cite{Damour:2014afa}. The latter situation is examined here in the case of two (non-spinning) black holes -in their center-of-mass reference system-  which start interacting at large
separation, approach each other, and finally separate again. 

The discussion here involves a two-body system with one of the two having a small  mass  (say $m_1$) with respect to the other (of mass $m_2$, $m_1\ll m_2$), so that
the backreaction of $m_1$ on the background metric generated by $m_2$ can be neglected.
The small body is assumed to deviate from geodesic motion in the background (Schwarzschild) gravitational field of the large mass $m_2$ because of self-interaction effects modeled by the presence of a drag force term in the equations of motion, with components proportional to the associated 4-velocity components (see, e.g., Ref. \cite{Gair:2010iv} and references therein). 
Dissipative effects arise on the non-geodesic motion of $m_1$ [$m_2$ is supposed at rest with respect to the chosen coordinate system], and hence $m_1$ undergoes a quasi-elastic scattering process, whose main features are discussed in comparison with the corresponding geodesic motion with the same initial conditions. 
Within this approximation, we will estimate the loss of energy and angular momentum during the scattering process, and investigate the dependence of both the scattering angle and impact parameter on the strength of the drag force itself. 

We will use geometrical units and conventionally assume that greek indices run from $0$ to $3$ whereas latin indices run from $1$ to $3$.

\section{Forced motion in a Schwarzschild field}

Let us consider a test particle with mass $m_1=m$ moving on the equatorial plane of a Schwarzschild black hole with mass $m_2=M$. The line element written in standard coordinates reads 
\begin{eqnarray}
ds^2&=& g_{\alpha\beta}dx^\alpha dx^\beta\nonumber\\
&=&-N^2 dt^2+N^{-2} dr^2+r^2(d\theta^2 +\sin^2\theta d\phi^2)\,,
\end{eqnarray}
with $N=\sqrt{1-\frac{2M}{r}}$ denoting the lapse function.
Let $U=U^\alpha \partial_\alpha$ be the particle's four velocity with $U^\alpha=dx^\alpha/d\tau\equiv x^\alpha{}'$ ($\tau$ being the proper time parameter, $U^\theta=0$)
and $a(U)$ its four acceleration 
\begin{eqnarray}
a(U)&=&\left( t''+\frac{2M}{r^2 N^2}t'r'\right) \partial_t \nonumber\\
&+&\left(r''-rN^2\phi'{}^2-\frac{M}{r^2N^2}r'{}^2+\frac{MN^2}{r^2}t'{}^2\right)\partial_r\nonumber\\
&+&\left(\phi''+\frac{2}{r} r'\phi'  \right)\partial_\phi\,. 
\end{eqnarray}
Let us assume that the particle undergoes a drag force $f(U)$ chosen so that its components in the plane of motion are proportional to the corresponding components of the four velocity itself, i.e., $f^r \propto U^r$ and $f^\phi \propto U^\phi$, while the temporal component follows from the orthogonality condition of $f(U)$ and $U$, $f(U)\cdot U=0$, namely
\beq
f(U)=f^t \partial_t -\lambda \left(U^r \partial_r+ U^\phi \partial_\phi\right)\,, 
\eeq
with $\lambda$ a dimensionless constant modeling the physics of the dragging and  $f^t$ given by
\beq
f^t=-\frac{\lambda}{N^2 t'}\left(r^2\phi'{}^2+\frac{r'{}^2}{N^2}\right)\,.
\eeq
The equations of motions 
$m a(U)=f(U)$ then reduce to 
\begin{eqnarray}
\label{final_sys}
&& t''+\frac{2M}{r^2 N^2}t'r'=-\frac{\sigma}{N^2 t'}\left(r^2\phi'{}^2+\frac{r'{}^2}{N^2}\right)\nonumber\\
&& r''-rN^2\phi'{}^2-\frac{M}{r^2N^2}r'{}^2+\frac{MN^2}{r^2}t'{}^2=-\sigma r'\nonumber\\
&& \phi''+\frac{2}{r} r'\phi'=-\sigma \phi'
\end{eqnarray}
where $\sigma =\lambda/m>0$ (with the dimensions of a length$^{-1}$).
The equations for $\phi$ and $r$ can be both reduced to first order equations, i.e., 
\begin{eqnarray}
\label{dphi_dtau_nongeo}
\phi' &=& \frac{L_-}{r^2}e^{-\sigma(\tau-\tau_-)}\,,\nonumber\\ 
\phi(\tau)&=& \phi_- +L_-\int_{\tau_-}^\tau \frac{e^{-\sigma(\tau-\tau_-)}}{r(\tau)^2} d\tau \,,
\end{eqnarray}
and (by using the normalization condition for $U$, $U\cdot U=-1$)
\begin{eqnarray}
r'{}^2&=&N^2 \left(N^2 t'{}^2 -\frac{L_-^2}{r^2}e^{-2\sigma(\tau-\tau_-)} -1  \right)\,.
\end{eqnarray}
The temporal equation (coupled to the  $r$ and $\phi$ equations) can be rewritten as
\beq
\frac{d}{d\tau}\left(N^2 t' e^{\sigma (\tau-\tau_-)}  \right)= \frac{ \sigma}{t'} e^{\sigma (\tau-\tau_-)}\,,
\eeq
that is
\begin{eqnarray}
t'&=&\frac{e^{-\sigma (\tau-\tau_-)}}{N^2}\left[E_-^2+2 \sigma  \int_{\tau_-}^\tau N^2  e^{2\sigma (\tau-\tau_-)} \, d\tau \right]^{1/2}
\equiv\frac{e^{-\sigma (\tau-\tau_-)}}{N^2}{\mathcal E}_-\,,
\end{eqnarray}
with
\beq
\label{cale_definition}
 {\mathcal E}_-(\tau)^2=E_-^2+ 2 \sigma  \int_{\tau_-}^\tau N^2  e^{2\sigma (\tau-\tau_-)} \, d\tau \,,
\eeq
$E_-={\mathcal E}_-(\tau_-)$ being an integration constant.
Substituting then into the radial equation we have formally
\beq
\frac{r'{}^2}{N^2}= e^{-2\sigma(\tau-\tau_-)} \left(\frac{{\mathcal E}_-^2}{N^2}-\frac{L_-^2}{r^2}\right) -1\,.
\eeq
It is useful to introduce the dimensionless inverse radial variable $u=M/r$, so that the previous equation becomes
\begin{eqnarray}
\label{du_dtau_nongeo}
\left(\frac{du}{d\tau} \right)^2 
&=&  u^4 \left[  (e^{-\sigma (\tau-\tau_-)}{\mathcal E}_-)^2
-(1-2u)(1+j_-^2 u^2 e^{-2\sigma (\tau-\tau_-)}) \right]\,,
\end{eqnarray}
where  $j_-=L_-/M$ denotes the dimensionless angular momentum per unit mass.
Finally, using the evolution equation for $\phi$ Eq. (\ref{du_dtau_nongeo}) reads
\beq
\label{dudphi2}
\left(\frac{du}{d\phi}\right)^2= \frac{{\mathcal E}_-^2-e^{2\sigma(\tau-\tau_-)}}{j_-^2}-u^2 (1-2u)  +  \frac{2u}{j_-^2}e^{2\sigma(\tau-\tau_-)}\,.
\eeq

\subsection{Scattering process}

Let the particle undergo a scattering process between the proper time values $\tau_-$ and $\tau_+>\tau_-$ (with $T\equiv\tau_+-\tau_-$) corresponding to an initial (\lq\lq in'') state $(\tau_-,E_-,j_-,\phi_-,u_-=0)$ and a final (\lq\lq up'') state $(\tau_+,E_+,j_+,\phi_+,u_+=0)$. Eventually, we may have $\tau_-\to -\infty$ and $\tau_+\to +\infty$, even if in the numerical integration of the orbits this situation is never achieved.

Eq. (\ref{dudphi2}) implies at the \lq\lq in'' state 
\beq
\label{dudphi2in}
\left(\frac{du}{d\phi}\right)^2_{-}= \frac{E_-^2-1}{j_-^2}\,,
\eeq
while at the \lq\lq up'' state
\beq
\label{dudphi2up}
\left(\frac{du}{d\phi}\right)^2_{+}= \frac{{\mathcal E}_-(\tau_+)^2-e^{2\sigma  T}}{j_-^2}\,.
\eeq
By comparing the previous equations with the corresponding geodesic equations at the \lq\lq up'' state, i.e., 
\beq
\label{dudphi2up_n}
\frac{M}{u^2}\frac{d\phi}{d\tau}= j_+\,,\qquad
\left(\frac{du}{d\phi}\right)^2_{+}= \frac{E_+^2-1}{j_+^2}\,,
\eeq
one obtains the following variation laws of both angular momentum (exact) and energy (approximated, since the integral in (\ref{cale_definition}) involves $u(\tau)$)
\beq
\label{Epjpdef}
j_+ =  e^{-\sigma T}j_-\,,\qquad
E_+  \approx   E_-+\frac{\sigma T}{E_-}\,,
\eeq
during the full scattering process.

Actually, from a numerical point of view the asymptotic states correspond to some nonzero values $u_\pm=10^{-n}$ (with $n$ fixed, e.g., $n=5$) and not to the theoretical values $u_\pm=0$; therefore, the duration  $T$ of the scattering process is necessarily finite and can be computed by using the numerical value of $\tau$ such that $u(\tau)$ reaches the fixed threshold.
For the same reason, the value $E_+$ given by Eq. (\ref{Epjpdef}) is overestimated (since the integral in Eq. (\ref{cale_definition}) is also overestimated). We will provide a better estimate for $E_+$ below.

Finally, introducing the new variable $v=(u-1/6)/2$ we find a \lq\lq canonical" form
\beq
\label{eq_v_gen}
\left(\frac{dv}{d\phi}\right)^2= 4v^3 -\tilde g_2v-\tilde g_3
\eeq
with
\begin{eqnarray}
\tilde g_2&=&-\frac{e^{2\sigma (\tau-\tau_-)}}{j_-^2}+\frac{1}{12}\,,\qquad
\tilde g_3 = \frac{1}{216}+\frac{e^{2\sigma (\tau-\tau_-)}}{6j_-^2}-\frac{{\mathcal E}_-^2}{4j_-^2}\,,
\end{eqnarray}
where $\tau$ is a function of $\phi$.
Clearly,  the presence of the term $e^{2\sigma(\tau-\tau_-)}$ complicates matters with respect to the geodesic case ($\sigma=0$), shortly recalled below.

\subsection{Reference geodesic}

When $\sigma=0$, Eq. (\ref{dudphi2}) reduces to
\beq
\label{dudphigeo}
\left(\frac{du}{d\phi}\right)^2
=\frac{E_-^2-1}{j_-^2} -u^2(1-2u) +\frac{2u}{j_-^2}\,,
\eeq
or, equivalently, 
\beq
\label{eq_v}
\left(\frac{dv}{d\phi}\right)^2= 4v^3 -g_2v-g_3\,,
\eeq
with $g_2=\tilde g_2(\tau_-)$ and $g_3=\tilde g_3(\tau_-)$, i.e.,
\beq
g_2=-\frac{1}{j_-^2}+\frac{1}{12}\,,\qquad g_3= \frac{1}{216}+\frac{1}{6j_-^2}-\frac{E_-^2}{4j_-^2}\,,
\eeq
both {\it constant} in this case. The solution of Eq. (\ref{eq_v}) is written in terms of the Weierstrass elliptic function ${\mathcal P}$ (see, e.g., \cite{Scharf:2011ii} and references therein)
\begin{eqnarray}
v(\phi)&=& {\mathcal P}(\phi-\phi_0; g_2,g_3)\,,
\end{eqnarray}
where, in general, $\phi_0=\phi_0(E_-,j_-)$ (or, equivalently, $\phi_0=\phi_0(g_2,g_3)$) is chosen so that
\beq
{\mathcal P}'(\phi_0; g_2,g_3)=0\,,
\eeq
i.e., $r'(\phi)|_{\phi=\phi_0}=0$ and $\phi_0$ corresponds to distance of minimum approach. [Recall that ${\mathcal P}$ is an even function of its argument whereas ${\mathcal P}'$ is odd.]
In addition,
\beq
\tau=\frac{1}{L_-}\int^\phi r^2 (\phi) d\phi\,.
\eeq

The energy and angular momentum (conserved in this geodesic case, so that $E_-=E_+$, $j_-=j_+$) are given by
\begin{eqnarray}
\label{Emjm}
E_-^2 &=& \frac{(p-2)^2-4e^2 }{p(p-3-e^2)}\,, \qquad
j_-^2 = \frac{p^2}{(p-3-e^2)}\,,
\end{eqnarray}
when parametrized in terms of the dimensionless semi-latus rectum $p$ and eccentricity $e$.
The distance of minimum approach can also be expressed in terms of $p$ and $e$ as 
\beq
r_{\rm min}=\frac{Mp}{1+e}\,,\quad 
u_{\rm min}=\frac{1+e}{p}\equiv(1+e)u_p\,,
\eeq
whereas the corresponding expression in terms of  $E_-$ and $j_-$ requires the solution of a cubic equation.
It is also standard to introduce the impact parameter $b$ as
\beq
\frac{b}{M}  =\frac{j_-}{\sqrt{E_-^2-1}}
=\frac{p^{3/2}}{\sqrt{(p-4)(e^2-1)}}\,.
\eeq

The asymptotic values $\phi_\pm$ follow from the solution of Eq. (\ref{dudphigeo}), which can be expressed in a closed analytic form \cite{Chandrasekhar:1985kt}. 
To this end, it is useful to introduce the polar parametrization of the orbit, i.e.,
\beq
u=u_p (1+e \cos \chi )\,,\qquad \frac{du}{d\tau}=-u_p e \sin \chi  \frac{d\chi }{d\tau} 
\eeq
where the relativistic anomaly  $\chi  \in[-\chi _{\rm(max)}, \chi _{\rm(max)}]$ for unbound orbits ($\chi=0$ corresponding to the perihelion passage) and
\beq
\chi _{\rm(max)}={\rm arccos}\left(-\frac{1}{e}  \right)\,.
\eeq
Expressing Eqs. (\ref{dphi_dtau_nongeo}) and (\ref{du_dtau_nongeo}) (for $\sigma=0$ and recalling Eq. (\ref{Emjm}) with $p=1/u_p$) in terms of $\chi $ implies 
\beq
\label{dchi_dtau_geo}
\frac{d\chi }{d\tau}= \frac{ (1+e \cos \chi   )^2 [p-6-2 e\cos \chi   ]^{1/2}}{M p^{3/2} (p-3-e ^2)^{1/2}}
\eeq
and
\beq
\frac{d\phi}{d\chi }=\frac{p^{1/2}}{(p-6-2 e\cos\chi )^{1/2}}\,.
\eeq
The latter equation can be integrated as
\beq
\phi(\chi ) =\frac{\Psi}{\sqrt{e u_p}}  \left[ K(\Psi)-F\left(\cos\frac{\chi}{2},\Psi\right)\right]\,,
\eeq
with $\phi(0)=0$ and 
\beq
\label{psi_def}
\Psi=2\sqrt{\frac{e   u_p  }{ 1-6 u_p+2 eu_p }} \,,
\eeq
where $K(k)$ and $F(\sin\varphi,k)$ are the complete and incomplete elliptic integrals of the first kind, respectively, defined by
\beq
K(k)=\int_0^{\frac{\pi}{2}}\frac{dx}{\sqrt{1-k^2\sin^2x}}\,,\qquad
F(\sin\varphi,k)=\int_0^{\varphi}\frac{dx}{\sqrt{1-k^2\sin^2x}}\,.
\eeq 
The associated deflection angle $\delta_{\rm geo}(u_p,e)=2\phi(\chi _{\rm(max)})-\pi$ then turns out to be
\begin{eqnarray}
\label{defl_angle}
\delta_{\rm geo}(u_p,e) &=& \frac{2\Psi}{\sqrt{e u_p}} \left[ K\left(\Psi \right)  
-  F \left(\sqrt{\frac{e-1}{2e}}
,\Psi \right)\right]
-\pi\,.
\end{eqnarray}
For example, with $p=20$ and $e=3/2$ (so that $E_-=1.03334$, $j_-=5.20756$ and $b_-=20M$) we get $\delta_{\rm geo}\approx2.59122$ (i.e., $148.46608$ deg).

\section{Observer-dependent analysis of the motion}

It is convenient to examine the features of motion from the point of view of a family of observers.
A natural choice involves observers whose world lines are aligned with the coordinate time lines with unit tangent vector
\beq
n=\frac{1}{N}\partial_t\,,\qquad n^\flat =-Ndt\,,
\eeq
(the symbol $\flat$ denotes the associated 1-form family of fields) which can be completed with a spatial triad 
\beq
e_{\hat r}=\frac{1}{\sqrt{g_{rr}}}\partial_r\,,\qquad
e_{\hat \theta}=\frac{1}{\sqrt{g_{\theta\theta}}}\partial_\theta\,,\qquad
e_{\hat \phi}=\frac{1}{\sqrt{g_{\phi\phi}}}\partial_\phi
\eeq
to form a spacetime orthonormal tetrad.
The observers $n$ can be used to decompose the particle's 4-velocity as \cite{Jantzen:1992rg,Bini:1997ea,Bini:1997eb}
\beq
U=\gamma(U,n)[n+\nu(U,n)]\,,\qquad \nu(U,n)=\nu(U,n)^{\hat a}e_{\hat a}\,,
\eeq
where $\gamma(U,n)=NU^t$, $\nu(U,n)^{\hat \theta}=0$ and
\beq
\nu(U,n)^{\hat r}=\frac{U^r}{N^2 U^t}\,,\quad
\nu(U,n)^{\hat \phi}=\frac{rU^\phi}{N U^t}\,,
\eeq
with inverse relations (shortening the notation for convenience)
\begin{eqnarray}
\label{u_vs_nu}
U^t&=& \frac{\gamma}{N}\,,\qquad
U^r = \gamma N \nu^{\hat r}\,,\qquad
U^\phi = \frac{\gamma}{r}\nu^{\hat \phi}\,.
\end{eqnarray}
It is also convenient to split magnitude ($||\nu(U,n)||$) and unit vector ($\hat \nu(U,n)$) of the spatial velocity
\beq
\nu(U,n)=||\nu(U,n)||\hat \nu(U,n)\,,
\eeq
abbreviated as $\nu(U,n)= \nu \hat \nu$.
Let us denote by
\beq
\bar U= \gamma (\nu n +\hat \nu)\,,\qquad \hat \nu=\sin \alpha e_{\hat r}+\cos \alpha e_{\hat \phi}
\eeq
the unit (spatial) vector orthogonal to $U$ in the orbital subspace, with
\beq
\cot \alpha =\frac{\hat \nu^{\hat \phi}}{\hat \nu^{\hat r}}=Nr \frac{d\phi}{dr}\,.
\eeq
Similarly,  the 4-acceleration can be decomposed as
\beq
a(U)=[a(U)^{\hat r}\nu_{\hat r}+a(U)^{\hat \phi}\nu_{\hat \phi}] n +a(U)^{\hat r}e_{\hat r}+a(U)^{\hat \phi}e_{\hat \phi}
\eeq
due to its orthogonality with $U$. 
Furthermore, since $U^r \partial_r+ U^\phi \partial_\phi\equiv P(n)U$, where $P(n)=g+n\otimes n$ projects orthogonally to $n$, the drag force can be written as
\beq
f(U)=-\lambda \nu \bar U\,.
\eeq
The full set of equations of motion results in
\begin{eqnarray}
\label{eqmoto}
\frac{d\nu}{d\tau}&=& -\sigma \frac{\nu}{\gamma^2} +\frac{(N^2-1)\sin \alpha}{2\gamma N r}\,, \nonumber\\
\frac{d\alpha}{d\tau}&=& \gamma\frac{2 N^2\nu^2+(N^2-1)}{2Nr\nu}  \cos \alpha \,,\nonumber\\
\frac{dr}{d\tau}&=& \gamma N \nu\sin \alpha\,,
\end{eqnarray}
with the equation for $\alpha$ (as well as those for the coordinates $t$, $r$, $\phi$) not depending explicitly on $\sigma$.
Note that the following quantity is a constant of motion
\beq
r\gamma\nu\cos\alpha\,e^{\sigma\tau}={\rm const}\,,
\eeq
and can be used, e.g., to eliminate $\alpha$.
As it is easy to check, the system (\ref{eqmoto}) does not admit non-trivial equilibrium solutions.  

Instead of $r$, one can use the dimensionless inverse radial variable $u=M/r$ and replace the last equation by
\beq
M\frac{du}{d\tau}= -\gamma N u^2 \nu\sin \alpha\,.
\eeq
The remaining equations are
\begin{eqnarray}
\label{eqmoto2}
\frac{dt}{d\tau}&=& \frac{\gamma}{N}\,,\qquad
\frac{d\phi}{d\tau}= \frac{\gamma\nu }{r}\cos \alpha\,.
\end{eqnarray}
The first of these equations in the geodesic case reduces to ${dt}/{d\tau}={E}/{N^2}$.
We then assume that at the end of the scattering process the quantity $\gamma N$ defines the outgoing energy $E_+/N^2$, leading to $E_+=\gamma_+ N_+$, i.e., with quantities evaluated at the threshold $u=u_+$.

In the simple flat space situation, i.e., the case $M=0$ ($N=1$)  we have the solution
\begin{eqnarray}
\gamma \nu &=& \gamma_0 \nu_0 e^{-\sigma \tau}\,,\qquad
\cos\alpha = \frac{r_0}{r}\,,\qquad
r^2= r_0^2\left[1 +K^2 (1-e^{-\sigma \tau})^2\right]\,,\nonumber\\
\phi &=&{\rm arctan}\left( \frac{e^{\sigma\tau}}{K}-(1-e^{\sigma \tau})K \right) -{\rm arctan}\left(\frac{1}{K}\right)\,,
\end{eqnarray}
where $K={\gamma_0  \nu_0 }/{\sigma r_0 }$, (having assumed $\alpha_0=0$ and $\phi_0=0$ without any loss of generality)
with the asymptotic limits (for $K\not =1$, corresponding to a $\phi=$ constant solution, not relevant for the present discussion)
\begin{eqnarray}
&& \nu \to 0\,,\qquad  
\cos \alpha \to \frac{1}{\sqrt{1+K^2}}\,,\qquad
r\to r_0\sqrt{1+K^2} \,,\nonumber\\
&& \phi \to \frac{\pi}{2}-{\rm arctan}\left(\frac{1}{K}\right)\,,
\end{eqnarray}
when $\tau \to \infty$, corresponding to the particle at rest.
In the generic case, instead, the above system can only be studied numerically.

\section{Numerical integration of the orbits}

Let us consider first the geodesic equations, with initial conditions taken at the distance of minimum approach corresponding to $r_0=r_{\rm min}=Mp/(1+e)$ and $\phi_0=0$, namely we fix the values of $p$ and $e$ which determine $E_-$ and $j_-$ from Eq. (\ref{Emjm}), and 
\beq
\nu_0=\frac{1+e}{\sqrt{p-2+2e}}\,,\qquad
\alpha_0=0\,.
\eeq
For example, taking the same values of $p=20$ and $e=1.5$ (as before) implies $r_0 = 8M$, $\nu_0 \approx 0.54554$, $E_-=1.03334$, $j_-=5.20756$, $\delta_{\rm geo}\approx2.59122$.
For numerical purposes, let us consider asymptotic radial states (theoretically corresponding to $u_\pm=0$) at finite $u_\pm=10^{-n}$ (with $n$ fixed, e.g., $n=5$). Therefore, the duration $T$ of the scattering process is necessarily finite and one can determine the values of $\tau_\pm$ such that $u(\tau_\pm)\le10^{-n}$, i.e., within the fixed threshold.
In particular, the above choice of parameters with $u(\tau_\pm)=10^{-5}$ implies $\tau_\pm\approx\pm3.83777\times10^5$, $\phi_\pm^{\rm(g)}\approx\pm2.86621$, $\alpha_\pm^{\rm(g)}\approx\mp1.57060$, $\nu^{\rm(g)}_\pm\approx0.25201$.
The \lq\lq in'' and \lq\lq up'' states for the geodesics are then fully specified. 

\begin{figure}
\centering
\[\begin{array}{cc}
\includegraphics[scale=0.3]{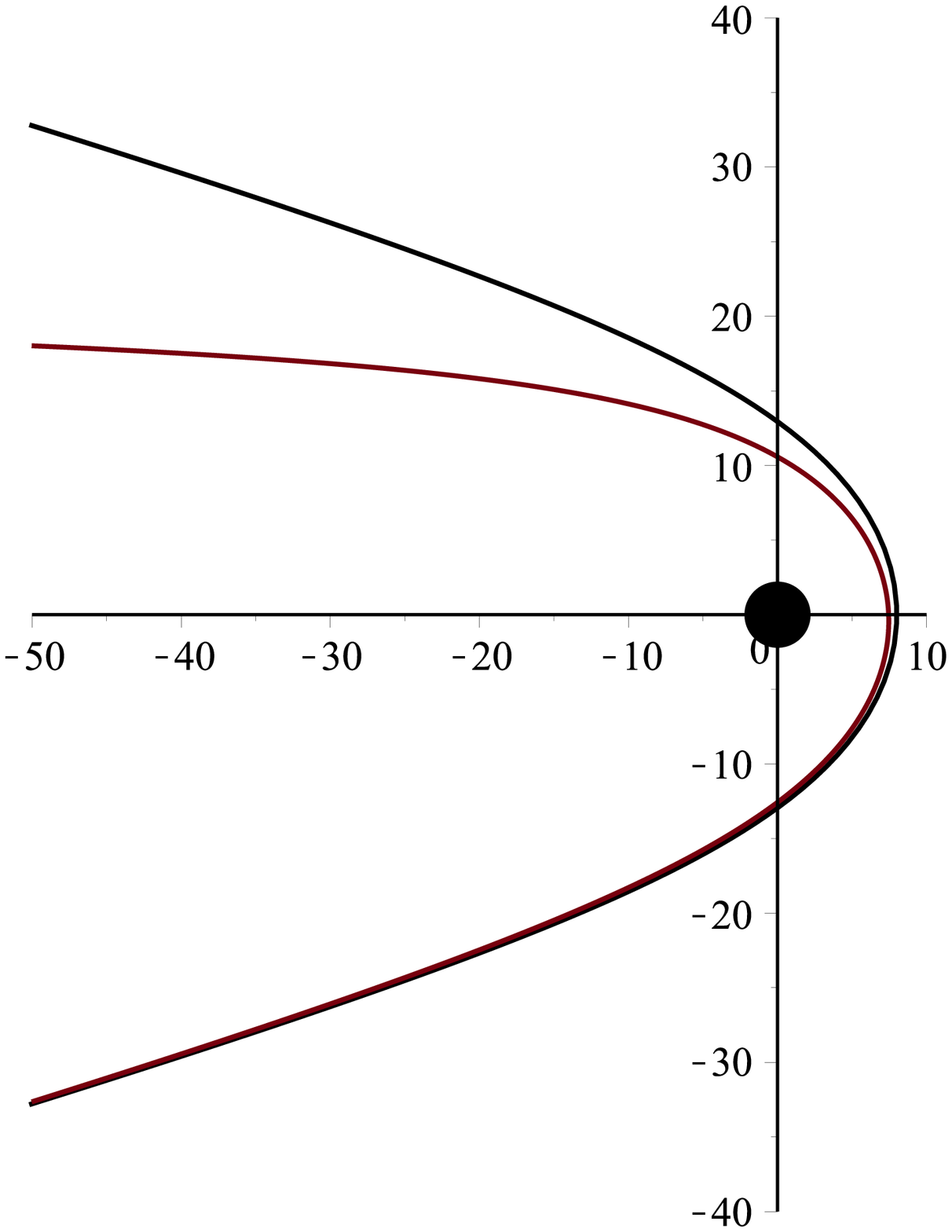}&\includegraphics[scale=0.3]{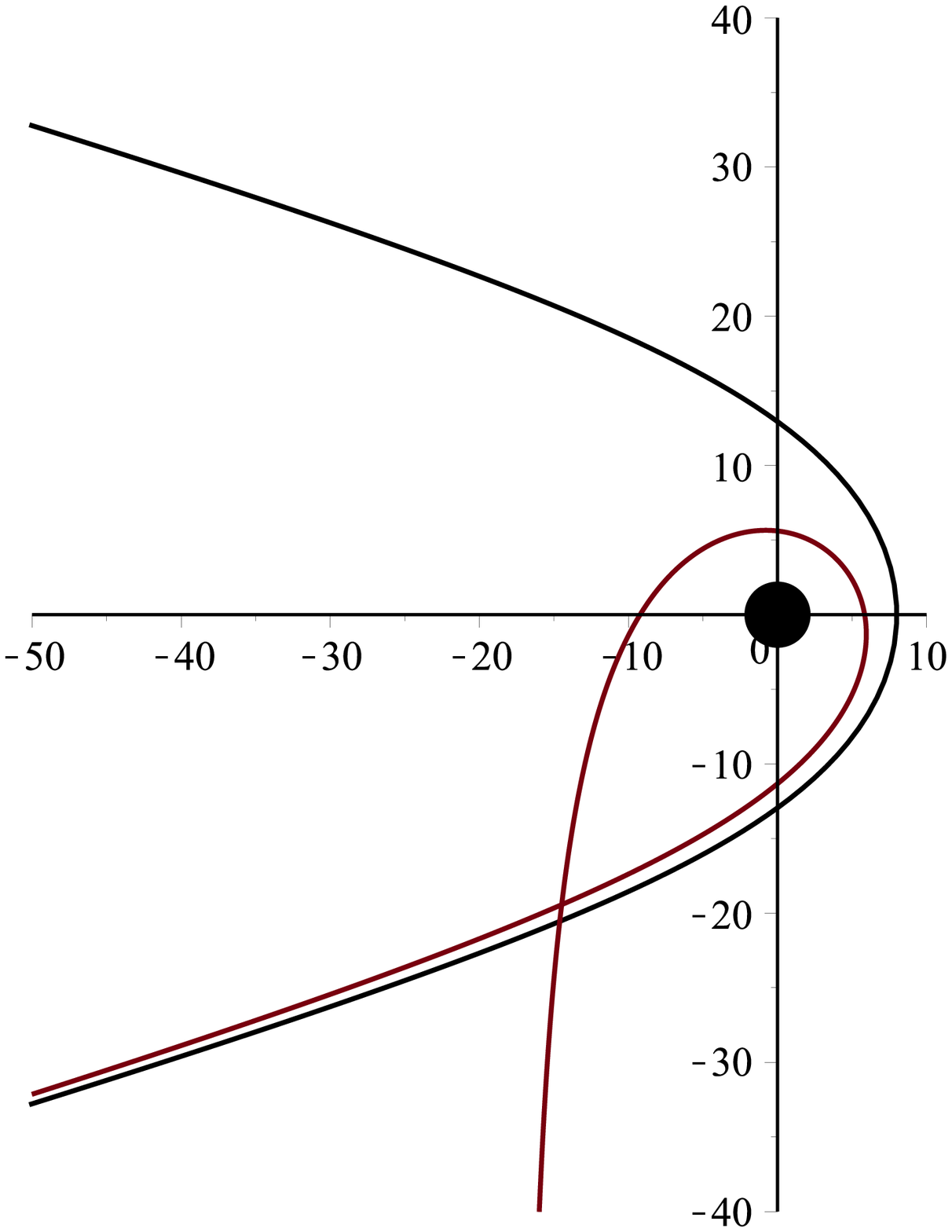}\\
(a)&(b)\\
\includegraphics[scale=0.3]{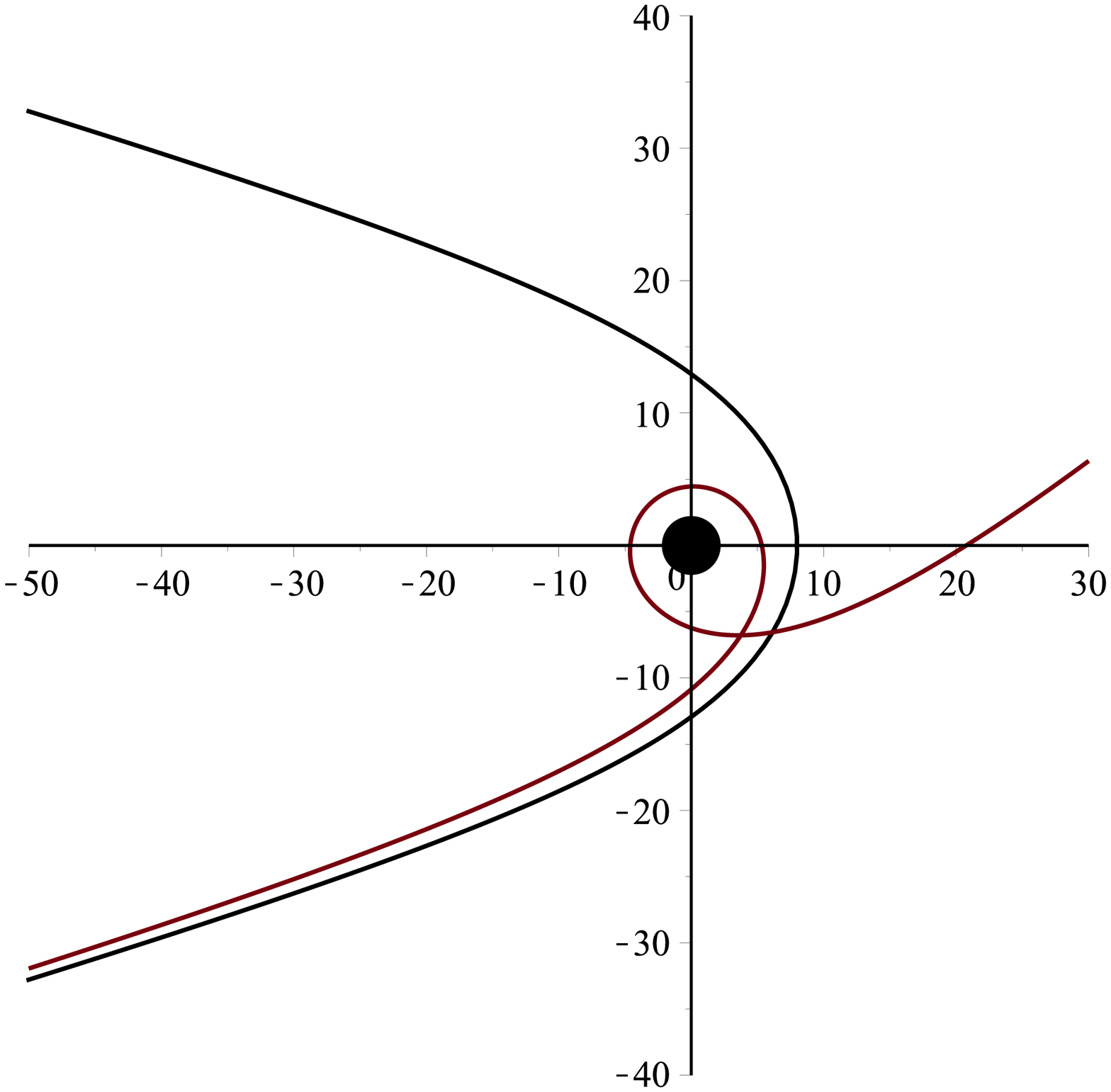}&\includegraphics[scale=0.3]{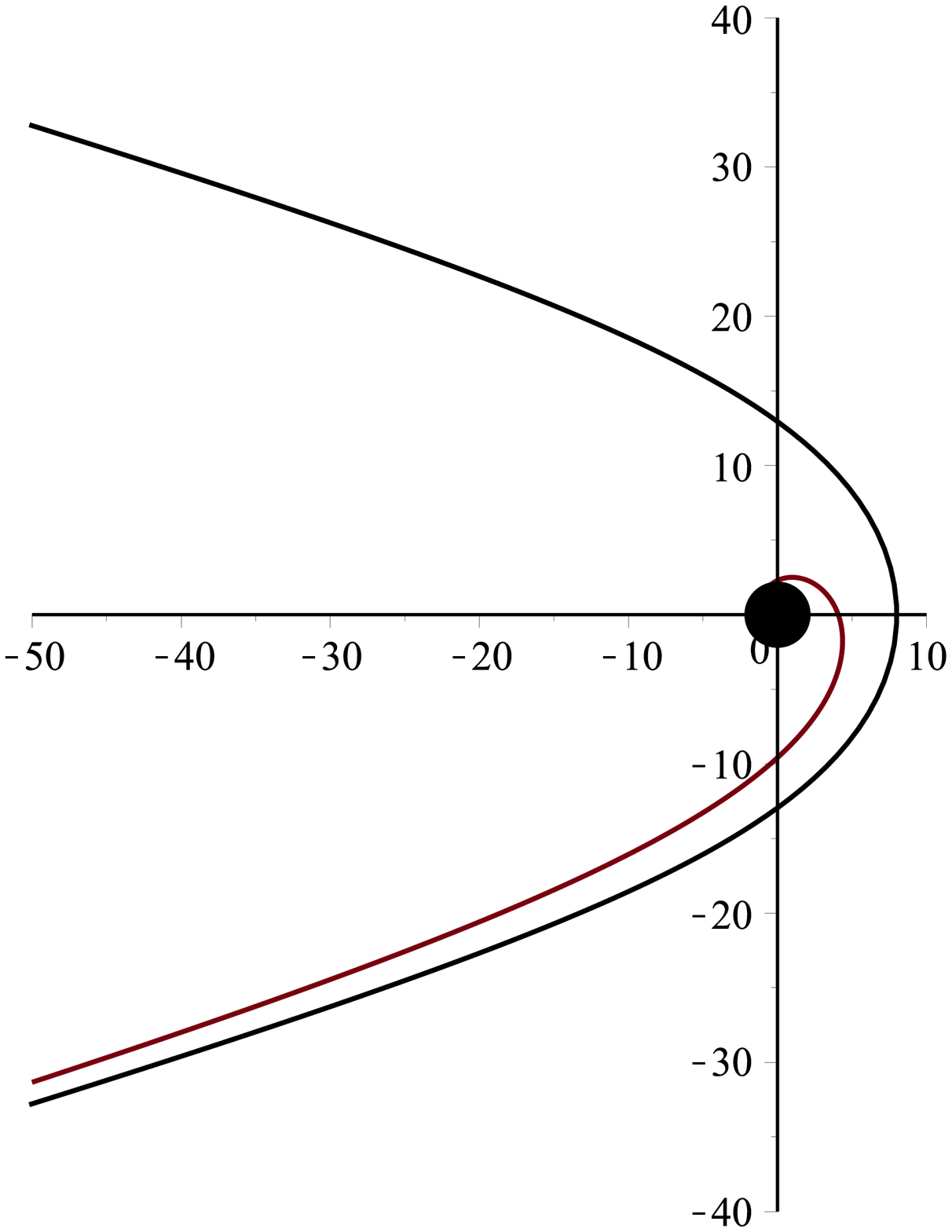}\\
(c)&(d)\\
\end{array}
\]
\caption{The geodesic (black, symmetric with respect to the horizontal axis) and accelerated (red) orbits are shown in the $r-\phi$ plane for the choice of parameters and initial conditions given in the text and different values of $\tilde\sigma=[1,4,5,7.5]\times10^{-7}$ (from (a) to (d)).
}
\label{fig:1}
\end{figure}

\begin{figure}
\centering
\[\begin{array}{cc}
\includegraphics[scale=0.25]{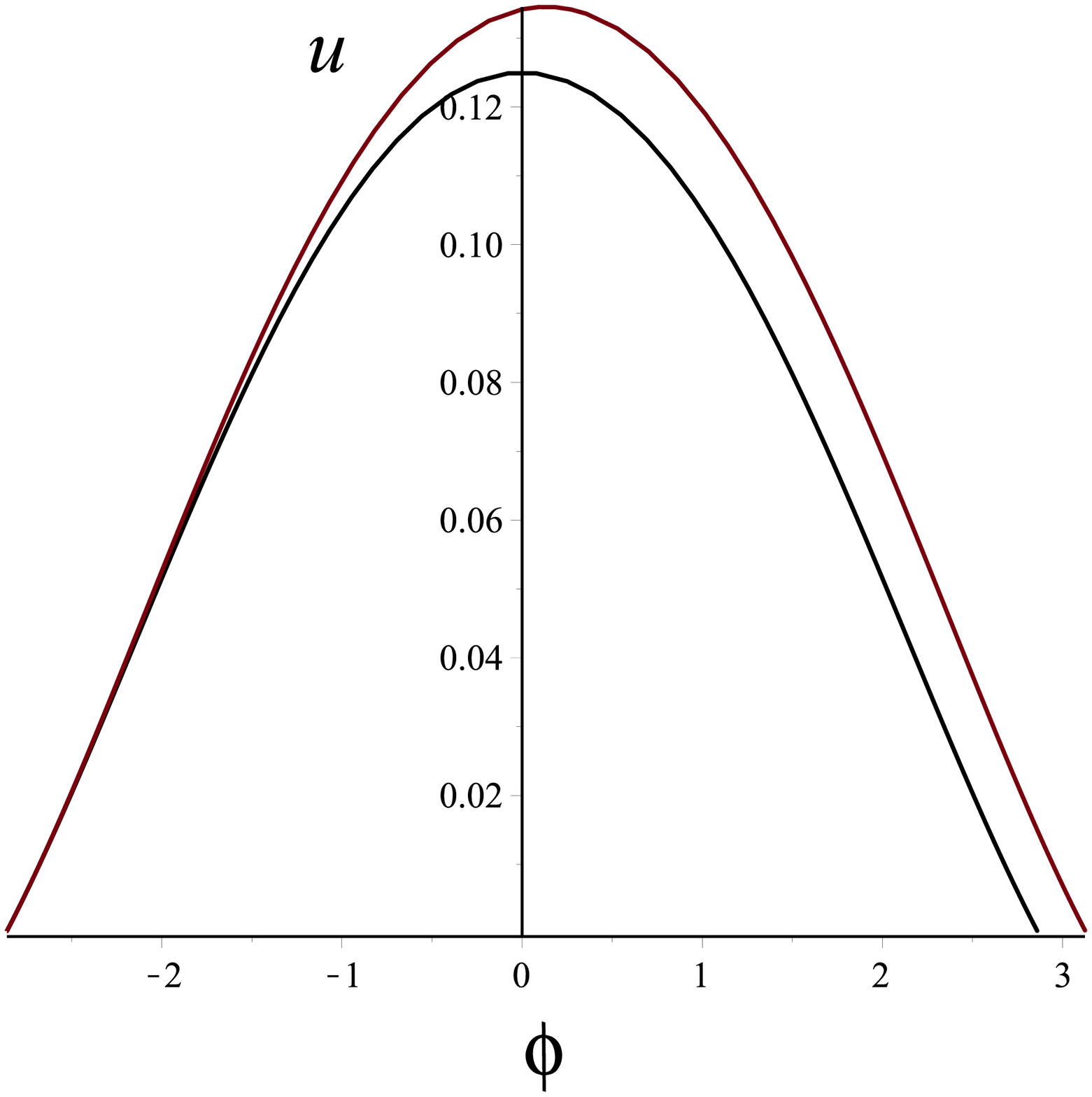}&\includegraphics[scale=0.25]{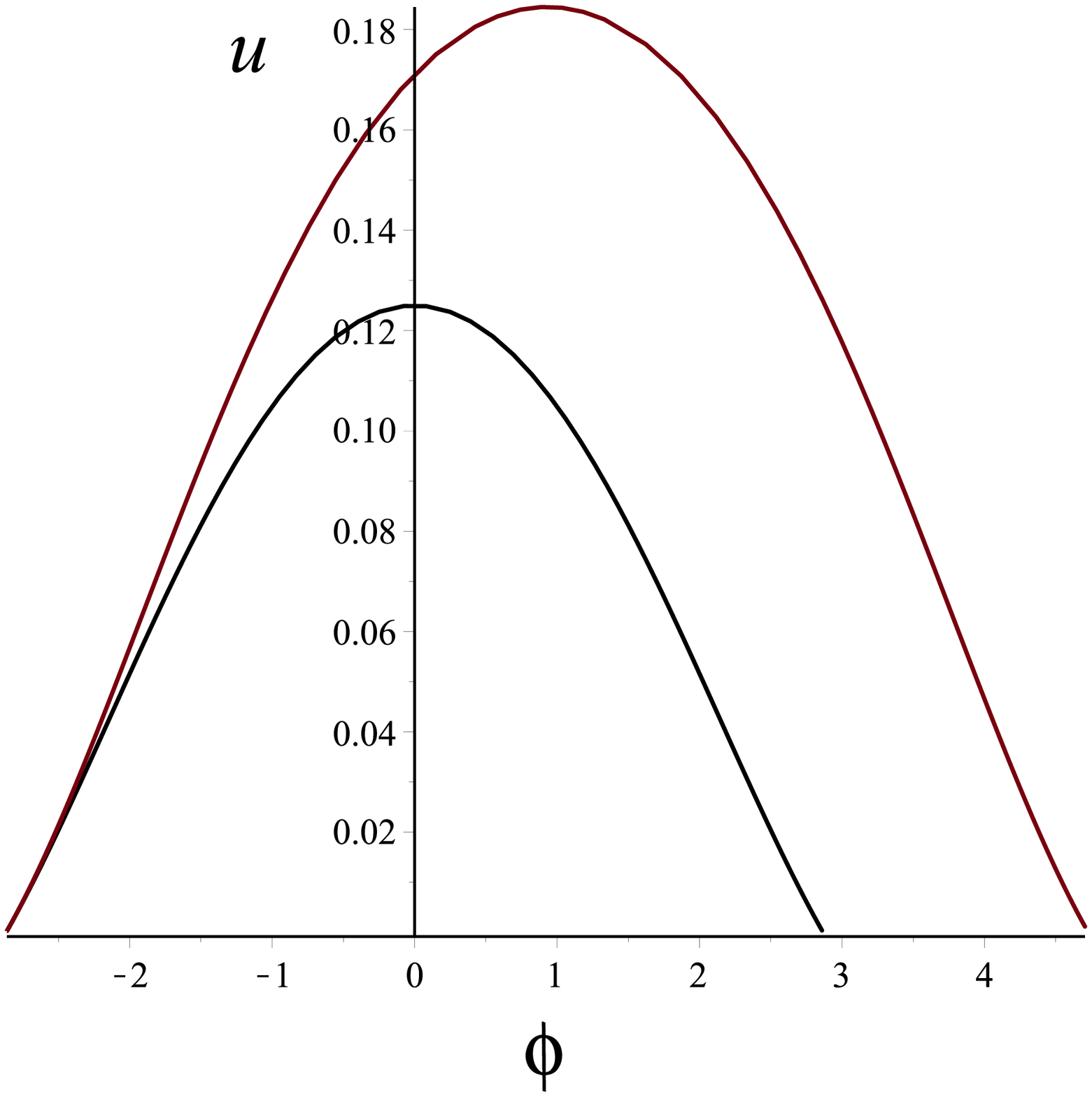}\\
(a)&(b)\\
\includegraphics[scale=0.25]{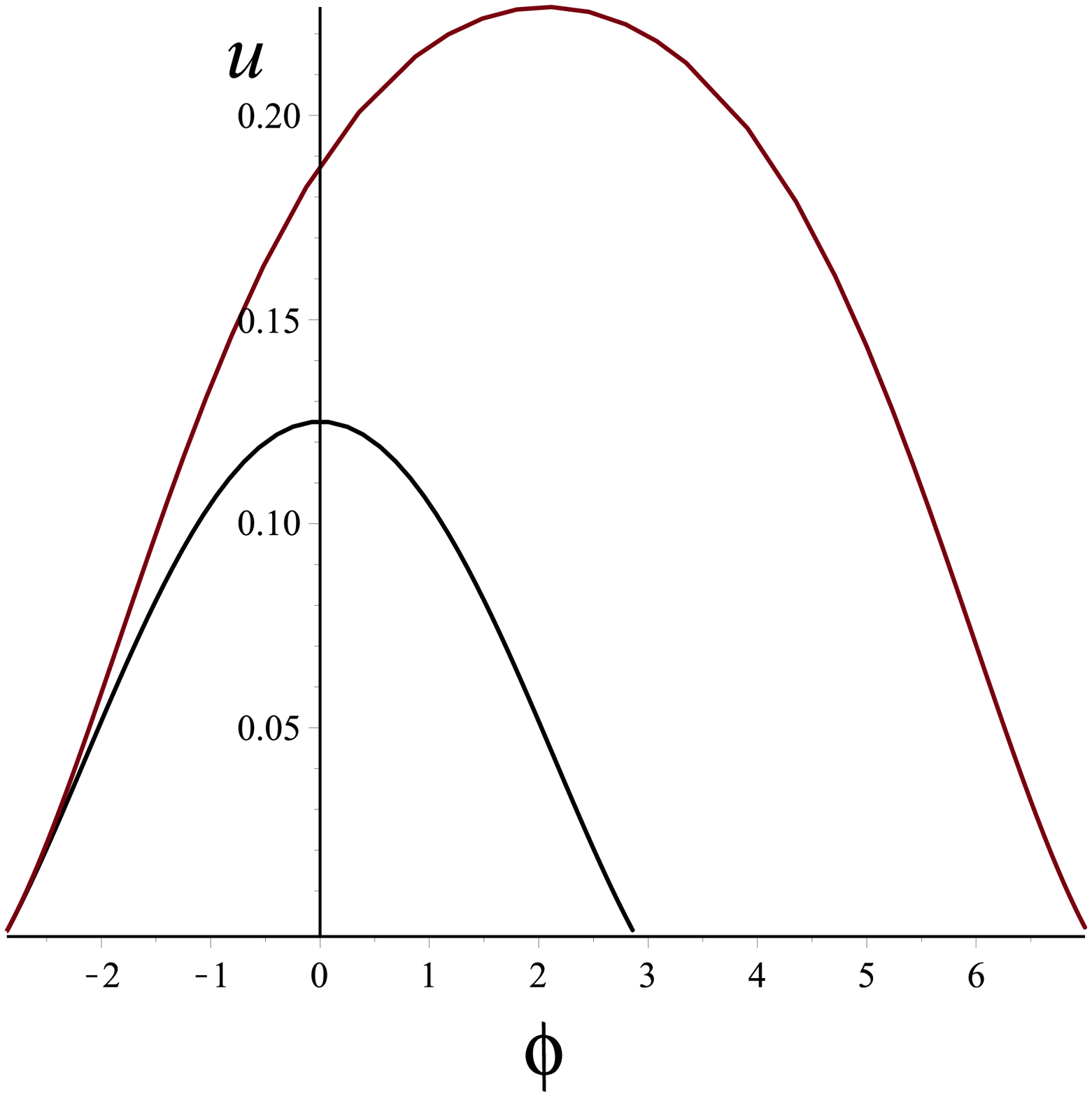}&\includegraphics[scale=0.25]{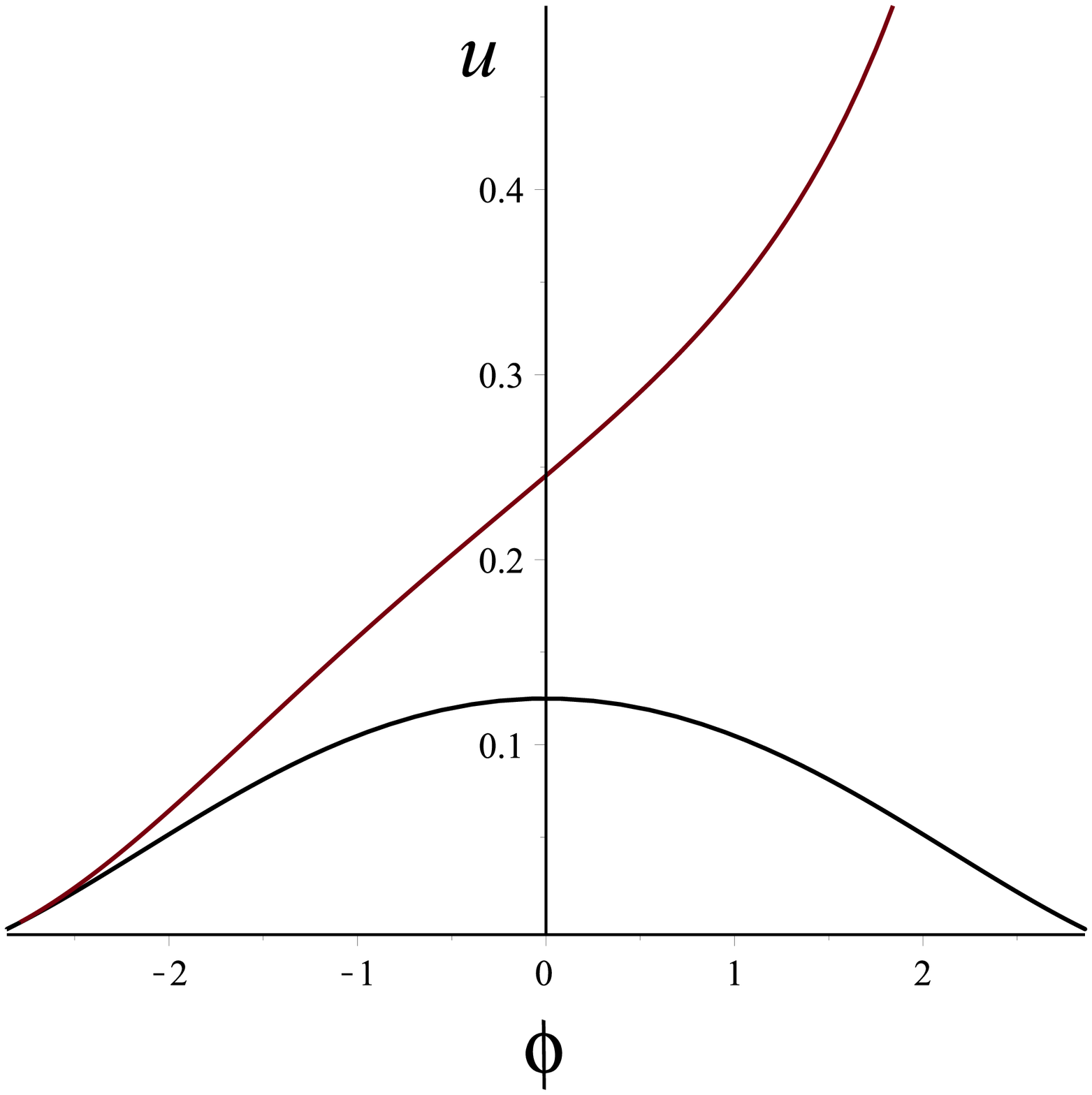}\\
(c)&(d)\\
\end{array}
\]
\caption{The behavior of $u(\phi)$ is shown for both the geodesic (with an even symmetry with respect to the $\phi$-axis,  $u(-\phi)=u(\phi)$) and accelerated orbits for the same choice of initial conditions as in Fig. \ref{fig:1} and the same values of $\tilde\sigma=[1,4,5,7.5]\times10^{-7}$ (from (a) to (d)).
}
\label{fig:2}
\end{figure}

Let us turn now to the non-geodesic case, and assume that the \lq\lq in'' state coincides with the geodesic one.
Therefore, at $\tau=\tau_-$ both the particle undergoing the drag force and that moving along a geodesic orbit start their motion with the same values of energy and angular momentum (or eccentricity and semi-latus rectum). We then integrate the motion equations (\ref{eqmoto}) and (\ref{eqmoto2}) with initial conditions $u(\tau_-)=10^{-5}$, $\phi(\tau_-)=\phi_-^{\rm(g)}$, $\alpha(\tau_-)=\alpha_-^{\rm(g)}$, $\nu(\tau_-)=\nu_-^{\rm(g)}$, for a fixed value of $\sigma$.
The deflection angle is now a function of $\sigma$ for fixed values of $p$ and $e$ and depends on the specified threshold for $u_\pm$.
The above choice of semi-latus rectum and eccentricity of the reference geodesic (leading to a distance of minimum approach close enough to the hole) implies that that very small values of the parameter $\tilde\sigma\equiv M\sigma$ are able to significantly modify the geodesic motion, causing strong deflections or even capture by the hole. 
The results are summarized in Figs. \ref{fig:1}--\ref{fig:2} and Tables \ref{tab:1}--\ref{tab:2}.

Figure \ref{fig:1} shows the trajectory of the particle undergoing drag force effects (with different strength $\tilde\sigma$) compared with the corresponding geodesic orbit. 
For increasing values of $\tilde\sigma$ the scattering angle increases as well (see Fig. \ref{fig:1} (a) to (c)) until the energy and angular momentum losses are enough to imply capture by the hole (see Fig. \ref{fig:1} (d)). The corresponding values of the scattering observables (scattering angle, energy and angular momentum losses) for different values of $\tilde\sigma$ are listed in Tables \ref{tab:1} and \ref{tab:2}.
It is interesting to point out that while the energy loss is rather small (fractionally some percent of the initial value), the angular momentum varies considerably during the scattering process for increasing $\tilde\sigma$. 

Figure \ref{fig:2} shows instead the full evolution of $u$ during the entire scattering process as a function of the azimuthal angle $\phi$. In the geodesic case the curve is symmetric about the vertical axis, whereas in the non-geodesic case this symmetry is lost, the value of the impact parameter slightly increasing with $\tilde\sigma$. 
The figures (see panels (a), (b) and (c)) also show the shift in the distance of minimum approach corresponding to the maximum of each curve, either geodesic or accelerated.

                          
\begin{table}
\centering
\caption{The results of numerical integration of the equations of motion for selected values of $\tilde \sigma$: angular deviation, energy, angular momentum, impact parameter and scattering angle.
}
\begin{tabular}{|l|l|l|l|l|l|}
\hline
$\tilde \sigma$& $\phi_+$ & $E_+$&   $j_+$ &   $b_+/M$ & $\delta\, ({\rm deg})$\cr 
\hline
$1.00\times10^{-8}$&2.89040&1.03285&5.16760&19.99753&149.82910\cr
$5.00\times10^{-8}$&2.99219&1.03087&5.00773&19.99915&155.66153\cr
$1.00\times10^{-7}$&3.13297&1.02849&4.80796&20.00138&163.72781\cr
$2.00\times10^{-7}$&3.47557&1.02399&4.40831&20.00672&183.35723\cr
$3.00\times10^{-7}$&3.95138&1.01986&4.00856&20.01347&210.61896\cr
$4.00\times10^{-7}$&4.72725&1.01611&3.60879&20.02213&255.07290\cr
$4.50\times10^{-7}$&5.42055&1.01438&3.40896&20.02744&294.79596\cr
$4.75\times10^{-7}$&5.98818&1.01355&3.30880&20.03045&327.31901\cr
$5.00\times10^{-7}$&7.01614&1.01275&3.20888&20.03386&386.21664\cr
\hline
\end{tabular}
\label{tab:1}
\end{table}


\begin{table}
\centering
\caption{The results of numerical integration of the equations of motion for selected values of $\tilde \sigma$: energy and angular momentum losses.
}
\begin{tabular}{|l|l|l|}
\hline
$\tilde \sigma$& $\Delta E$&   $\Delta j$ \cr 
\hline
$1.00\times10^{-8}$&-4.93642$\times10^{-4}$&-0.03996\cr
$5.00\times10^{-8}$&-2.46961$\times10^{-3}$&-0.19982\cr
$1.00\times10^{-7}$&-4.85658$\times10^{-3}$&-0.39960\cr
$2.00\times10^{-7}$&-9.35491$\times10^{-3}$&-0.79924\cr
$3.00\times10^{-7}$&-1.34810$\times10^{-2}$&-1.19899\cr
$4.00\times10^{-7}$&-1.72290$\times10^{-2}$&-1.59876\cr
$4.50\times10^{-7}$&-1.89594$\times10^{-2}$&-1.79859\cr
$4.75\times10^{-7}$&-1.97907$\times10^{-2}$&-1.89876\cr
$5.00\times10^{-7}$&-2.05960$\times10^{-2}$&-1.99868\cr
\hline
\end{tabular}
\label{tab:2}
\end{table}

\section{Conclusions and outlook}

We have considered corrections to the hyperbolic motion around a Schwarz\-schild black hole due to a drag force, which is a realistic situation in many astrophysical processes.
For instance, this is the case of particles interacting with accreting flows also in the presence of external electromagnetic fields or plasma \cite{McCourt:2015dpa}, not examined in the present study.
The drag force used here acts on the orbital plane like a viscous force with components proportional to the particle 4-velocity components by a dimensionless constant parameter $\tilde \sigma$ representing the strength of the (self) interaction, following previous studies (see Appendix D of Ref. \cite{Gair:2010iv}).
Besides discussing the general features of the motion, we have performed several numerical analyses to study the loss of energy and angular momentum of the particle undergoing such a drag effect as well as the dependence of the scattering angle and impact parameter on $\tilde \sigma$. The results are summarized in Figs. \ref{fig:1}--\ref{fig:2} and Tables \ref{tab:1}--\ref{tab:2}. We have focused on strong field effects, i.e., the distance of minimum approach has been chosen close enough to the horizon of the hole. We have found that significant effects arise even in the case of very small values of $\tilde \sigma\sim10^{-7}-10^{-8}$. Indeed, a fine tuning of $\tilde \sigma$ in this range implies capture by the hole, or scattering at large angles. In the latter case, the scattering angle has been computed by evolving the particle's trajectory in comparison with the corresponding geodesic orbit emanating from a common initial condition.  
Our analysis shows that the chosen form of the drag force field only allows for scattering or capture. This is an interesting situation, especially in comparison with other kinds of dragging effects, like the Poynting-Robertson effect \cite{PR}, where the presence of a superposed photon test field implies the existence of \lq\lq suspended'' (equilibrium) orbits, as discussed in recent literature \cite{abram,ML,Bini:2008vk,Bini:2011zza}. 

The present analysis is preliminary and complementary to a forthcoming study which goes beyond the test particle approximation by using self-force techniques and standard perturbation theory to consistently generate the drag force as entirely due to the particle's emission of gravitational radiation \cite{Martel:2003jj}.

\begin{acknowledgements}
D.B. thanks the Italian INFN (Naples) and  ICRANet for partial support. 
\end{acknowledgements}


\begin{thebibliography}{99}  


\bibitem{ligo} 
See the website of the LIGO Scientific Collaboration at http://www.ligo.caltech.edu/.

\bibitem{virgo} 
See the website of the VIRGO Scientific Collaboration at http://www.virgo.infn.it/. 


\bibitem{Abbott:2016blz} 
  B.~P.~Abbott {\it et al.} [LIGO Scientific and Virgo Collaborations],
  ``Observation of Gravitational Waves from a Binary Black Hole Merger,''
  Phys.\ Rev.\ Lett.\  {\bf 116}, no. 6, 061102 (2016)
  [arXiv:1602.03837 [gr-qc]].
 

\bibitem{Pretorius:2007jn} 
  F.~Pretorius and D.~Khurana,
  ``Black hole mergers and unstable circular orbits,''
  Class.\ Quant.\ Grav.\  {\bf 24}, S83 (2007)
  [gr-qc/0702084]. 

\bibitem{Shibata:2008rq} 
  M.~Shibata, H.~Okawa and T.~Yamamoto,
   ``High-velocity collision of two black holes,''
  Phys.\ Rev.\ D {\bf 78}, 101501 (2008)
  [arXiv:0810.4735 [gr-qc]].
 
\bibitem{Sperhake:2008ga} 
  U.~Sperhake, V.~Cardoso, F.~Pretorius, E.~Berti and J.~A.~Gonzalez,
   ``The High-energy collision of two black holes,''
  Phys.\ Rev.\ Lett.\  {\bf 101}, 161101 (2008)
  [arXiv:0806.1738 [gr-qc]].
 
\bibitem{Sperhake:2009jz} 
  U.~Sperhake, V.~Cardoso, F.~Pretorius, E.~Berti, T.~Hinderer and N.~Yunes,
   ``Cross section, final spin and zoom-whirl behavior in high-energy black hole collisions,''
  Phys.\ Rev.\ Lett.\  {\bf 103}, 131102 (2009)
  [arXiv:0907.1252 [gr-qc]].

\bibitem{Witek:2010xi} 
  H.~Witek, M.~Zilhao, L.~Gualtieri, V.~Cardoso, C.~Herdeiro, A.~Nerozzi and U.~Sperhake,
   ``Numerical relativity for D dimensional space-times: head-on collisions of black holes and gravitational wave extraction,''
  Phys.\ Rev.\ D {\bf 82}, 104014 (2010)
  [arXiv:1006.3081 [gr-qc]].

\bibitem{Sperhake:2012me} 
  U.~Sperhake, E.~Berti, V.~Cardoso and F.~Pretorius,
   ``Universality, maximum radiation and absorption in high-energy collisions of black holes with spin,''
  Phys.\ Rev.\ Lett.\  {\bf 111}, no. 4, 041101 (2013)
  [arXiv:1211.6114 [gr-qc]].

\bibitem{Damour:2014afa} 
  T.~Damour, F.~Guercilena, I.~Hinder, S.~Hopper, A.~Nagar and L.~Rezzolla,
   ``Strong-Field Scattering of Two Black Holes: Numerics Versus Analytics,''
  Phys.\ Rev.\ D {\bf 89}, no. 8, 081503 (2014)
  [arXiv:1402.7307 [gr-qc]].

\bibitem{Gair:2010iv} 
  J.~R.~Gair, E.~E.~Flanagan, S.~Drasco, T.~Hinderer and S.~Babak,
   ``Forced motion near black holes,''
  Phys.\ Rev.\ D {\bf 83}, 044037 (2011)
  [arXiv:1012.5111 [gr-qc]].
 
\bibitem{Scharf:2011ii} 
  G.~Scharf,
  ``Schwarzschild geodesics in terms of elliptic functions and the related red shift,''
  J.\ Mod.\ Phys.\  {\bf 2}, 274 (2011)
  [arXiv:1101.1207 [astro-ph.GA]].
  
\bibitem{Chandrasekhar:1985kt} 
  S.~Chandrasekhar,
  ``The mathematical theory of black holes,''
  OXFORD, UK: CLARENDON (1985) 646 P.
 
\bibitem{Jantzen:1992rg} 
  R.~T.~Jantzen, P.~Carini and D.~Bini,
  ``The Many faces of gravitoelectromagnetism,''
  Annals Phys.\  {\bf 215}, 1 (1992)
  [gr-qc/0106043].

\bibitem{Bini:1997ea} 
  D.~Bini, P.~Carini and R.~T.~Jantzen,
  ``The Intrinsic derivative and centrifugal forces in general relativity. 1. Theoretical foundations,''
  Int.\ J.\ Mod.\ Phys.\ D {\bf 6}, 1 (1997)
  [gr-qc/0106013].

\bibitem{Bini:1997eb} 
  D.~Bini, P.~Carini and R.~T.~Jantzen,
  ``The Intrinsic derivative and centrifugal forces in general relativity. 2. Applications to circular orbits in some familiar stationary axisymmetric space-times,''
  Int.\ J.\ Mod.\ Phys.\ D {\bf 6}, 143 (1997)
  [gr-qc/0106014].
 
\bibitem{McCourt:2015dpa} 
  M.~McCourt and A.~M.~Madigan,
  ``Going with the flow: using gas clouds to probe the accretion flow feeding Sgr A*,''
  Mon.\ Not.\ Roy.\ Astron.\ Soc.\  {\bf 455}, no. 2, 2187 (2016)
  [arXiv:1503.04801 [astro-ph.HE]].

\bibitem{PR} 
J.~H.~Poynting,
``Radiation in the Solar System: Its Effect on Temperature and Its Pressure on Small Bodies,''
Phil. Trans. Roy. Soc. {\bf 202}, 525 (1904);
H.~P.~Robertson,
``Dynamical effects of radiation in the solar system,''
MNRAS {\bf 97}, 423 (1937)

\bibitem{abram}
M.~A.~Abramowicz, G.~F.~R.~Ellis, and A.~Lanza, 
``Relativistic effects in superluminal jets and neutron star winds,''
Astrophys. J. {\bf 361}, 470 (1990)

\bibitem{ML} 
M.~C.~Miller and F.~K.~Lamb,
``Motion of Accreting Matter near Luminous Slowly Rotating Relativistic Stars,''
Astrophys. J. {\bf 470}, 1033 (1996)

\bibitem{Bini:2008vk} 
  D.~Bini, R.~T.~Jantzen and L.~Stella,
  ``The General relativistic Poynting-Robertson effect,''
  Class.\ Quant.\ Grav.\  {\bf 26}, 055009 (2009)
  [arXiv:0808.1083 [gr-qc]].
 
\bibitem{Bini:2011zza} 
  D.~Bini, A.~Geralico, R.~T.~Jantzen, O.~Semerák and L.~Stella,
  ``The general relativistic Poynting-Robertson effect: II. A photon flux with nonzero angular momentum,''
  Class.\ Quant.\ Grav.\  {\bf 28}, 035008 (2011)
  [arXiv:1408.4945 [gr-qc]].

\bibitem{Martel:2003jj} 
  K.~Martel,
  ``Gravitational wave forms from a point particle orbiting a Schwarzschild black hole,''
  Phys.\ Rev.\ D {\bf 69}, 044025 (2004)
  [gr-qc/0311017].

    
\end{thebibliography}
\end{document}